\documentstyle[titlepage,12pt]{article}
\begin{document}
\parindent 2em
\baselineskip 4.5ex

\begin{titlepage}
\begin{center}
\vspace{12mm}
{\LARGE  The anomalous diffusion in high magnetic field and the
quasiparticle density of states}
\vspace{25mm}

Igor F. Herbut and Zlatko Te\v sanovi\' c \\
Department of Physics, Johns Hopkins University, Baltimore, MD 21218

\end{center}
\vspace{10mm}

\noindent
{\bf Abstract:} We consider a disordered two-dimensional electronic
system in the limit of high magnetic field at the metal-insulator
transition. Density of states close to the Fermi level acquires a divergent
correction to the lowest order in
electron-electron interaction and shows a new power-law dependence
on the energy, with the power given by the anomalous diffusion
exponent $\eta$. This should be observable in the tunneling experiment
with double-well GaAs heterostructure
of the mobility $\sim 10^{4}cm^{2}V/s$, at fields of $\sim 10T$
at temperatures of $\sim 10mK$ and voltages of $\sim 1 \mu V$.


\end{titlepage}

\section{Introduction}

One-parameter scaling theory of localization predicts that, at
least in the absence of time-reversal breaking fields and interactions,
there is a single independent critical exponent, the one for the
localization length, which characterizes the
metal-insulator transition as a critical point {\cite 1}.
In particular, the exponent which governs the diffusion behavior
and the eigenstates correlations at
the mobility edge is set by one-parameter scaling hypothesis to
$\eta=2-d$, where $d$ is the dimensionality of the system.
In strong magnetic field however, the observation of integer quantum
Hall effect signals a breakdown of one-parameter scaling in
two-dimensional (2d) electron system {\cite 2}.
Within each disorder-broadened
Landau level (LL) there is an unique value of the energy where the
extended states reside and the system undergoes a metal-insulator
transition when the Fermi level is at the mobility edge.
Chalker and co-workers {\cite 3} demonstrated another surprising
feature of the localization transition in strong fields:
right at the mobility edge the diffusion constant
develops a dependence on the frequency
and the wave-vector, $D(q^{2}/\omega)\propto(q^{2}/\omega)^{-\eta/2}$ for
large $q^{2}/\omega$, with a {\it non-zero}
value of the exponent $\eta$. Equivalently,
the wave-functions
at the mobility edge  are neither truly extended nor localized,
but instead are fractals with the generalized dimension
$d_{eff}=2-\eta$ {\cite 4} {\cite 5}.  The value of the exponent $\eta$ is
expected to be an universal quantity which characterizes the strong-field
limit and independent of the details of a disorder potential {\cite 6}.

The result that the exponent $\eta$ is non-trivial in high magnetic
field has been conjectured from an exact inequality
satisfied by the two-particle spectral function in the lowest LL
and by using the hypothesis of scale-invariance at the mobility edge {\cite 3}.
Subsequently, the conclusion
has been confirmed in numerical calculations {\cite 3},{\cite 5},{\cite 6}.
Experimentally however, the anomalous diffusion exponent
has yet to be observed,
and in that context it is important to find measurable consequences
of this new scaling behavior at the mobility edge. Shimshoni and Sondhi
{\cite 7} have argued that the temperature dependence of Coulomb drag
between two electron layers at the transition yields the information
on the exponent $\eta$.
Brandes et al. {\cite 8} have
suggested that $\eta$ should appear in the power of the
temperature dependence of the energy loss rate of hot
electrons due to phonon emission and of the quasiparticle lifetime due
to electron-phonon interaction. The quasiparticle
lifetime determined by electron-electron interactions however, is not
altered by the anomalous diffusion exponent in any significant way
{\cite 8}, {\cite 9}.
In this paper we demonstrate that when the Fermi level is at
the mobility edge, Coulomb interactions between electrons in the
lowest LL strongly suppress
the quasiparticle density of states (DOS) close to the Fermi level.
In certain range of energies DOS now exhibits a
power-law dependence on the energy (at T=0) or on the temperature
(at $T\neq 0$) with power $\eta/2$, and our perturbative calculation
suggests that right at the Fermi level DOS vanishes. This implies a weak
zero-bias anomaly in the I-V characteristics of the tunneling into the
electronic layer which would provide a direct information on the anomalous
 diffusion exponent. We discuss the experimental conditions
under which this effect could be observed.

\section{Quasiparticle density of states}

 Consider a 2d electron layer in a magnetic field strong enough so
that all electrons are polarized and in the lowest LL. The
LL mixing due to the disorder potential or interactions is neglected
and we work in units in which the magnetic length $l=(\hbar c/eB)^{1/2}$
 is unity.
If $\psi_{n}(\vec{r})$ are the  lowest LL eigen-functions of a particular
realization of the disorder, the Hamiltonian in this basis is
given by
\begin{equation}
H=\sum_{n}E_{n}a_{n}^{+}a_{n}+ \frac{1}{2}
\sum_{n,m,p,q}\langle m,n|v_{c}|q,p\rangle a_{m}^{+} a_{n}^{+}
a_{p} a_{q},
\end{equation}
where $v_{c}=e^{2}/r$ is the Coulomb interaction and we measure
energy from the Fermi level. First we
assume $T=0$, and calculate the self-energy correction to the single-particle
Green's function $G_{m}(\omega)=(\omega-E_{m}-\Sigma_{m}(\omega))^{-1}$
within the perturbation theory. The
real part of the disorder-averaged self-energy at energy $E$,
\begin{equation}
\Sigma_{E}(\omega)=N_{0}^{-1}(E) \langle \sum_{m} \delta(E-E_{m})
\Sigma_{m}(\omega)\rangle,
\end{equation}
determines to the lowest order in the interaction, the quasiparticle
DOS via \cite{10}:
\begin{equation}
N(E)=N_{0}(E)(1+\frac{d}{dE} Re\Sigma_{E}(E))^{-1}.
\end{equation}
$N_{0}(E)$ is the average DOS of non-interacting electrons
in the lowest LL and in the random potential and angular brackets
denote the averaging over disorder.
Consider first the
exchange contribution to the self-energy:
\begin{equation}
\Sigma_{E}^{exc}(\omega)=-\frac{1}{N_{0}(E)} \int_{-\infty}^{0}dE'
 \frac{d^{2}\vec{q}}{(2\pi)^{2}} S(E,E',q) v_{scr}(\omega,q),
\end{equation}
where the disorder-averaged two-particle spectral density is defined by
\cite{10}
\begin{equation}
S(E,E',q)\equiv \int d^{2}\vec{r} \exp{ (i\vec{q}\vec{r})}
\langle \sum_{n,m} \delta(E-E_{n})\delta(E'-E_{m})\psi_{n}^{*}(\vec{r})
\psi_{m}^{*}(0) \psi_{m}(\vec{r}) \psi_{n}(0) \rangle.
\end{equation}
For small momentum and energies close to the mobility edge
the two-particle spectral density takes the
 familiar diffusive form {\cite 3}, \cite{10}:
\begin{equation}
S(\nu,q)=\frac{N_{0}(0)}{\pi}
\frac{D(\nu,q)q^{2}}{\nu^{2}+D^{2}(\nu,q)q^{4}},
\end{equation}
and we assume the
generalized diffusion constant $D(\nu,q)=\hbar D f(q^{2}/
cN_{0}(0)\nu)$, $\nu=|E-E'|$,
$c$ and $D$ are constants and $f(y)=1$ for $y<1$ and
$f(y)=y^{-\eta/2}$ for $y>1$ \cite{3}.
 It is essential for our discussion that even though we are calculating
the lowest order contribution to
the average self-energy we include the effect of
{\it screening} on Coulomb interaction in eq. 4.
By definition, the dynamically screened Coulomb interaction is given by:
\begin{equation}
v_{scr}(\omega,q)= v_{c}(q)(1+v_{c}(q)\Pi(\omega,q))^{-1}
\end{equation}
and we assume that the polarization function has it's standard  RPA form:
\begin{equation}
\Pi(\omega,q)=\frac{N_{0}(0) D(\omega,q)q^{2}}{-i\omega+D(\omega,q)q^{2}}.
\end{equation}
As will be shown shortly, this assumption is by no means essential for
our results. To calculate the correction to the quasiparticle DOS,
 we will need only the static limit of the screened Coulomb interaction.
Thus, as long as the polarization function has a diffusive form, the
effects beyond simple RPA approximation which would be represented by
a more complicated diffusion constant, exactly cancel out in the
calculation.
Important point is that the static part of the
screened interaction is short ranged. Using the equations 4,6,7,8 after
some calculation we find, for $E\approx 0$:
\begin{equation}
\frac{d}{dE} Re\Sigma^{exc}_{E} (E)=
\frac{F_{exc}(x,E)}{2\pi g}(\frac{|E|}{\Delta})^{-\frac{\eta}
{2}},
\end{equation}
where the function $F_{exc}(x,E)$ is given by an integral
\begin{equation}
F_{exc}(x,E)=x \int_{(|E|/\Delta)^{1/2}}^{1} \frac{dt}{t^{1-\eta}(t+x)}.
\end{equation}
$g=\sigma_{xx}/(e^{2}/h)$ is the dimensionless dissipative
conductance $\sigma_{xx}=e^{2}DN_{0}(0)$,
the energy scale is $\Delta=\pi \Gamma/cg$,
$\Gamma=2\hbar D$ is the half-width of the disorder-broadened  LL,
and $x=(e^{2}/\epsilon d)/\Gamma$, where $d$ is the average distance
between electrons. The result in eqs. 9 and 10 is the same as if we
used the screened interaction in static approximation
in our calculation. Thus for the present purposes,
we could have set $\omega=0$ from the
beginning in the polarization function 8 so that
complications related to the form
of the diffusion constant in that expression cancel out.
Let us now analyze the equations 9 and 10.
 If $x<<1$, then for $x^{2}<<|E|/\Delta<<1$ we have
$d Re\Sigma^{exc}_{E}(E)/dE \approx (x/2\pi g(1-\eta)) (|E|/\Delta)^{-1/2}$
 (assuming $\eta<1$), just like the result would be
if the bare Coulomb interaction was used in the calculation.
However,
closer to the Fermi level, i.e.,
$|E|/\Delta\approx x^{2}$ and smaller,
one obtains the novel power-law dependence
on the energy: $d Re\Sigma^{exc}_{E}(E)/dE \approx (1/2\pi g \eta)
(|E|/\Delta)^{-\eta/2}$.

If one would perform the previous calculation using the delta-function
interaction instead of $v_{scr}(\omega,q)$ in eq.4, close to the Fermi level
self-energy would again diverge with the power
$\eta/2$, only with a different prefactor.  This means
that the same term must also exist in the Hartree contribution to the
self-energy, since the two should cancel each other if the interaction
is infinitely short ranged \cite{10,11} and electrons are completely
spin-polarized.
Taking both  the exchange and the direct
contributions into account, we finally obtain
that at zero temperature, sufficiently close to the Fermi level
the correction to DOS reads:
\begin{equation}
\frac{\delta N(E)}{N_{0}(0)}=\frac{F(x)}{ 2 \pi g \eta }
(\frac{|E|}{\Delta})^{-\frac{\eta}{2}}
\end{equation}
where $F(x)=\eta F_{exc}(x,0)-F_{dir}(x)$, and
\cite{12,13} $F_{dir}=N_{0}(0)\int d^{2}\vec{q}
v_{scr}(0,q)/\pi$. The last integral gives
\begin{equation}
F_{dir}(x)=2x-2x^{2}\ln(1+x^{-1}),
\end{equation}
and $x$ is the ratio between the interaction and disorder energy
scales as defined earlier.

Since we found that at $T=0$ frequency dependence of the screened
interaction does not affect the behavior of DOS sufficiently close to the Fermi
level, at finite temperatures we may neglect it completely. The
simple calculation then yields the result:
\begin{equation}
\frac{\delta N(E,T)}{N_{0}(0)}=\frac{F(x)}{2\pi \eta g}
(\frac{T}{\Delta})^{-\frac{\eta}{2}}
f(\frac{|E|}{T})
\end {equation}
for $|E|,T\approx 0$, where
\begin{equation}
f(z)=\int_{-\infty}^{+\infty}\frac{e^{t+z}}{|t|^{\frac{\eta}{2}}
(e^{t+z}+1)^{2}} dt .
\end{equation}
The function $f(z)=z^{-\eta/2}$ for $z>>1$ and $f(0)=0.92$ (for
$\eta=0.5$). Thus,
at zero temperature we recover the result in eq. 11. At $T\neq 0$,
DOS acquires
a power-law dependence on the temperature instead on the energy.

A few remarks are in order at this point.
The obtained power-law should
be compared to the logarithmically divergent correction
to DOS in $d\geq 2$ implied by one-parameter scaling
relation $\eta=2-d$ \cite{14}. The exponent $\eta$ is determined
completely by the geometry of the extended state and is not in any
sense a small parameter in the problem. The found power-law
is therefore a truly distinct behavior from the standard logarithmic
correction to DOS. This should be contrasted with the situations
where the exponent itself is a small quantity so that the expansion
in powers of this exponent
to the leading order agrees with the logarithmic behavior
(for a similar scenario see for instance ref. 15).
Also, the assumption that the Fermi level is right at the
mobility edge so that the
diffusion constant is a function of the combination $q^{2}/\omega$
and not only of the momentum is important. If this was not
so, i. e. if the Fermi level was slightly off but close to the
mobility edge, the power law in eq. 11 would be cut off below a finite energy
defined by $\xi(0)\approx D E_{cut}^{-1/2}$,
where $\xi(E)$ is the
localization length at energy $E$ ($E=0$ still defines the Fermi level).
For the frequencies smaller than
the $E_{cut}$ diffusion constant becomes a function of combination
$(q\xi(0))$, and independent of frequency.
With $\eta$ larger than zero  the perturbative correction of the DOS
then would show a
crossover from the power law divergence as in eq. 11 for $E>E_{cut}$ to
the standard logarithmic divergence for $E<E_{cut}$.

 The divergence of the first-order correction to DOS in our
problem indicates the
breakdown of the perturbation theory very close to the Fermi energy.
Since from the eq. 10 we saw that the power $\eta/2$ turns on
for $(|E|/\Delta) \approx x^{2}$ and below, one might wonder whether that is in
the region where higher order terms are already significant.
The point where the first order correction becomes of order unity
marks the region of energies
where are our lowest order perturbation theory becomes
insufficient: this happens at $(|E|/\Delta)\approx (F(x)/2\pi\eta g)^{2/\eta}$.
The function $F(x)$ behaves like $x^{\eta}/(1-\eta)$ for $x<0.01$,
peaks at $x\approx 0.2$ with the value of 0.26, and roughly stays constant
for $0.2<x<1$. Consider first the physically more relevant regime $0.2<x<1$:
then $x^{2}\approx 0.1$
and the perturbation theory breaks down for $E/\Delta\approx 10^{-3}<<x^{2}$.
Here we assumed $g\approx 0.5$ and $\eta\approx 0.5$ \cite{3,5,6}. The
important point is that there is a considerable range of energies
for which our simple lowest order perturbation theory is valid and the
result is indeed given by eqs. 11 and 13 (see Figure 1). One may also
consider the regime when  $x\rightarrow 0$: in that
case the breakdown occurs for $|E|/\Delta \approx x^{2}$, so that again
right before it occurs the non-trivial power law appears, although the
result is not as clear-cut as for larger $x$. Even though we can
not say with certainty what happens very close to the Fermi level
where the problem becomes non-perturbative, we still
expect however that the perturbation theory is qualitatively correct
and that DOS will indeed vanish right at the Fermi level. Standard
scaling arguments would suggest that ultimately
DOS still goes to zero as a power-law, but with the exponent determined
by the full interacting theory. Recent numerical calculations which
include Coulomb interactions at the Hartree-Fock level find, quite
surprisingly, both $\eta$ and the localization length
exponent to be the same as
in the non-interacting problem \cite{16}. This points to the
intriguing possibility
that the exponent which ultimately determines the behavior of the DOS
at the Fermi level is not different from the one we calculated. This issue
deserves more attention in the future.

\section{Tunneling}

The obtained correction of the quasiparticle DOS will modify the
thermodynamic quantities when the Fermi level is at the
mobility edge. For instance, the specific heat will acquire
a low-temperature correction: $\delta C_{v}/T\propto T^{\eta/2}$.
However, thermodynamic quantities are difficult to measure since
a typical 2d electron layer in GaAs heterostructure contains very few
($\sim 10^{11}$) particles. Recently, tunneling experiments have
proved  to be a new useful tool for studying 2d electron
systems \cite{17,18,19}. In what follows, we therefore
 consider  a typical tunneling experimental setup consisting
of two identical, parallel 2d electron layers separated
by a semiconductor barrier. When a small voltage is applied between the
layers, the tunneling current is:
\begin{equation}
I\propto N^{2}(eV,T) V
\end{equation}
with the constant of proportionality being roughly the inverse
of the tunneling resistance for non-interacting electrons.
When the Fermi levels in both layers are  tuned to the mobility edge,
the tunneling conductance should exhibit a power-law anomaly as a
function of the voltage ($T<<eV$) or of the
temperature ($eV<<T$) with power $\eta$.
Let us now estimate the relevant temperature (or voltage) scale over which
the anomaly should become observable. DOS will start
to show a power-law behavior when the correction becomes
comparable to unity. This yields
the temperature scale $T\approx (\pi\Gamma/
cg)(F (x)/2\pi\eta g)^{2/\eta}$ as already discussed. We assume
the numbers for $g\approx 0.5 $ and $\eta\approx 0.5 $ as before.
The value of the constant $c$ is disorder dependent and less well known.
Here, we take $c\approx 60$, which was
found for the white-noise random potential {\cite 3}. All the numbers put
together give the relevant temperatures to be $T\propto 10^{-4}*\Gamma$.
Note that according to our discussion, this is deep
in the region where the power is determined by the screened Couloumb
interaction.

In a typical experiment,  GaAs samples of very
high mobility (for example, $\mu=2*10^{6}cm^{2}V/s$ in ref. 18) are used.
{}From the zero-field mobility
we can estimate the LL broadening in the following way: assume that the
scattering in the system is caused by the potential $V(\vec{r})=\lambda\sum_{i}
\delta(\vec{r}-\vec{r_{i}})$, where \{$\vec{r_{i}}$\} are 2d
coordinates of uniformly distributed scatterers with the density
$n_{imp}$. The zero-field
mobility is $\mu=e\tau_{0}/m$, $\tau_{0}=
\hbar/(n_{imp}\lambda^{2}\pi N_{0})$ is the scattering
time  in Born approximation and
$N_{0}=2m\pi/h^{2}$ is 2d electron DOS in zero magnetic field.
The band mass in GaAs is $m=0.07m_{e}$.
In strong magnetic field, Born approximation gives the
lowest LL broadening in the same potential $\Gamma\approx
(\lambda^{2}n_{imp}/l^{2})^{1/2}$.
At the magnetic field of $10T$ one then
obtains $\Gamma\approx 6K$, an order of
magnitude smaller than the Coulomb energy
scale which is around 50K (at $10T$, $\hbar\omega_{c}\approx 160K$).
We treated the interactions as a perturbation to the disorder
problem, so we need the LL broadening to be at least of the same
order of magnitude as the Coulomb energy scale to be in the
regime where our calculation is applicable. Presented estimate suggests that
$\Gamma$ is proportional to the inverse of the square root of the
mobility, hence decreasing the mobility of the sample by a factor of 100
would make $\Gamma\approx 60K$, somewhat larger than the Coulomb energy
but still sufficiently smaller than the cyclotron energy so that
neglecting LL mixing is still a reasonable approximation. The
temperature when the tunneling anomaly should occur is
then $\sim 0.01K$, or in
terms of the voltage around $1\mu V$.

\section{Conclusion}

To summarize, we demonstrated using the
perturbation theory that the interaction effects in
2d electron system in high magnetic field at the
metal-insulator transition cause a diverging power-law correction of the
quasi-particle density of states close to the Fermi level.
This is interpreted as a precursor of the power-law vanishing
of the density od states at the Fermi energy. The power in the perturbative
regime is
given by the anomalous diffusion exponent, so a tunneling
experiments may be used to test the novel diffusion behavior in
high magnetic field discussed by Chalker. An estimate
of the energy scales involved shows that if a double-well
GaAs heterostructure of zero-field
mobility of the order of $10^{4}cm^{2}V/s$ is used in the tunneling
experiment at fields of $10T$,
a zero-bias anomaly  in the
current-voltage characteristics should
develop for temperatures of the order of $10mK$ and voltages
$\sim 1 \mu V$.

\pagebreak

Captions:

Figure 1. Schematic view at the behavior of the quasiparticle
density of states for small energies with the Fermi level at the mobility
edge. The value of the small parameter $x^{2}\approx 0.1$
marks the crossover point from the the regime where
the behavior is essentially determined by the bare
Coulomb interaction (power 1/2)
 into the regime where screening becomes effective
and determines the non-trivial power law (power $\eta/2$). In the shaded region
the problem becomes non-perturbative.

\pagebreak

\end{document}